# Resource Letter MDS-1: Mobile Devices and Sensors for Physics Teaching

Martín Monteiro
Universidad ORT Uruguay, Uruguay
and
Arturo C. Martí
Instituto de Física, Universidad de la República, Iguá 4225, Montevideo, 11200, Uruguay
monteiro@ort.edu.uy and marti@fisica.edu.uy

**ABSTRACT**

This Resource Letter provides a guide to the literature on teaching experimental physics using sensors in tablets, smartphones, and some specialized devices. After a general discussion of the hardware (sensors) and the software (apps), we present resources for experiments using mobile-device sensors in many areas of physics education: mechanics, oscillations and waves, optics, electromagnetism, matter, modern physics, and astronomy.

## I. INTRODUCTION

In recent years there has been a revolution in the use of cell phones. In the late 2000s, smartphones appeared in the market and quickly became extremely popular: about $1.5 \times 10^9$ smartphones were sold in 2020, and the vast majority of physics students around the world have access to a smartphone.

Smartphones can be a useful experimental tool not only because they have a relatively powerful processor and can access wireless networks, but also due to the presence of a number of built-in sensors (see Table 1 below). Sensors can determine the orientation of the device, disable the touch screen, adjust the display's brightness according to external conditions to save the battery, geolocate the device, monitor the user's heart rate, and unlock the device. The set of sensors included in mobile devices large and is expected to expand in the coming years.

The relevance of the sensors goes far beyond those everyday uses mentioned: in the past few years numerous educational and scientific applications have been proposed, including physics experiments. As the original intention of sensor manufacturers was not to use these sensors in science or physics experiments, some effort is needed to transform smartphones into scientific instruments. However, the reward for making this effort is considerable. Even mid-range smartphones incorporate a broad range of sensors that would otherwise be difficult to acquire. Since most students already have such a device, they can be combined with easily available equipment and materials to create a low-cost, compact, multi-sensor experimental platform with no need for complicated connections or setup. Another factor, less objective, is that students do not consider a smartphone to be a strange or magical device, since it is part of their daily lives. Whether or not this fact influences the learning process is still under examination.

We might compare smartphones to devices designed specifically for the use of instructional laboratory experiments in physics and science, including the SmartCart® (Pasco), PocketLab® (Myriad Sensors, Inc), IOLab® (MacMillan), and Go Direct® Sensor Cart (Vernier). These devices share some capabilities with smartphones, but do not offer the traditional phone functions. In these devices, the selection of sensors, the connectivity and the software have been designed for pedagogy, with student-friendly interfaces. However, these devices are still alien to the student's experience, as compared to a smartphone. In addition, these devices must be purchased specifically for the physics laboratory, as students will not already own them. There also exist several platforms, such as Arduino or BBC mibro:bit (see Refs. [5, 12, 14, 15, 16] below), that have been adapted to perform experiments using sensors. Unlike smartphones in which the set of sensors is predetermined by the manufacturers, these programmable boards allow one to choose from a wide range of sensors and modules including ultrasound sensors, flame detectors, infra-red emitters and receptors, water level sensors, relays, PIR (Passive infra-red), and metal detectors. We have included a few articles using Arduino boards as examples of experiments that could be accomplished using a combination of Arduino boards and smartphones.

Communication with sensors requires specific software, known as an *app,* specially designed to be executed by the mobile device. Since most of the mobile devices currently used are based on two operating systems, iOS and Android, we will discuss the most useful apps running in these operating systems, but many others also may be useful in specific circumstances.

This Resource Letter is aimed at physicists and scientists interested in adapting or performing experiments using mobile-device sensors to collect *real* physical data. Other potential uses of the mobile devices in the classroom such as accessing the internet, sharing data among students, interacting with teachers, clicker-like activities, simulations, among many others, are not considered here. Video analysis of physical phenomena, which has received a lot of attention in the last decades, is also not included in this Resource Letter. In the selection of references, preference has been given to papers in which the experiments can be performed using state-of-the-art software and hardware. Within the selection criteria we prioritized articles published in journals—but we also cite some of the few books already published—that present experiments with sufficient detail and clarity that they will be straightforward to implement in a classroom.

The organization of this Resource Letter is as follows. As the topic is relatively new, there are only a few books which are cited in Section II. After that, resources about hardware and useful apps (software) are discussed. Next, we focus on experiments, starting with references covering multiple fields topics and then focusing on specific fields (mechanics, oscillations and waves, optics, electromagnetism, matter, modern physics, and astronomy). Finally, we discuss citizen science, educational proposals, and impact in Physics Education Research (PER).

## II. BASIC RESOURCES

### A. Journals and special issues

The recent literature addressing physics experiments using smartphone and mobile-device sensors in physics is substantial. Since 2011, hundreds of journal papers have been published. Most of the articles cited in this Resource Letter have been published in the following journals:

*American Journal of Physics:* https://aapt.scitation.org/journal/ajp

*European Journal of Physics,* https://iopscience.iop.org/journal/0143-0807

*Physics Education,* https://iopscience.iop.org/journal/0031-9120

*Revista Brasileira de Ensino de Física,* http://www.sbfisica.org.br/rbef/

*The Physics Teacher,* https://aapt.scitation.org/journal/pte

It is worth highlighting that the regular column, *iPhysicsLab*, has been published monthly in *The Physics Teacher* since 2012. Also, the *Papers in Physics* journal published a Focus Series on *Low-cost Experiments in Physics using recent technologies as sensors or open source hardware,* located at https://www.papersinphysics.org/papersinphysics/lowcostexperimentsinphysics. Another mini review worth mentioning was published by J. Stoop, *New ways to use smartphones for science, J. Stoop (Elsevier Connect, 2017)* located at: https://www.elsevier.com/connect/new-ways-to-use-smartphones-for-science. This virtual special issue from Elsevier is a must-read ensemble of papers previously published in journals. It includes resources in citizen science relatively accessible to the public, covering topics such as environment, health, space, smart cities, and society. Emphasis is given to experiments or measurements that can be conducted by the public following collaborative or crowd-sourcing methodologies.

## B. Books, eBooks, and internet resources

The following references are suitable choices for an introductory approach to the use of smartphone sensor in the physics laboratories.

1. **Kinematic Labs with Mobile Devices**, J. M. Kinser (Morgan & Claypool Publishers, 2015). DOI: https://doi.org/10.1088/978-1-6270-5628-1. This book (also available as an eBook) presents a comprehensible introduction to the use of smartphone sensors by first-year students at home or in places where laboratories are not easily accessible. It presents 13 experiments requiring only a smartphone and widely available equipment and also includes a brief overview of typical sensors and data analysis using typical apps and PC software. (I)

2. **iStage 2: Smartphones in Science Teaching,** edited by U. Hänsler, S. Schlunk and J. Schulze (Science on Stage Deutschland e.V., Germany, 2014). Published by the Science on Stage Europe group (https://www.science-on-stage.eu/), it offers an introduction to the use of smartphones and a dozen simple experiments aimed at secondary-school teachers in mathematics, physics, chemistry, and biology. A free download available in German and English from this link: https://www.science-on-stage.eu/smartphones. (E)

3. **Mobile Device Models**, W. Christian, C. Countryman, F. Esquembre. This internet website https://www.compadre.org/books/?ID=49&FID=45990 hosted by the AAPT project ComPADRE, emerged from the popular computer-generated *Physlet* (**Phy**sics content simulated with Java app**lets**) animations, and provides *Easy Java* simulations, problems and lectures designed to demonstrate physical concepts involving sensors and mobile devices. It is especially devoted to the use of the accelerometer for experiments with inclines, pendulums, and rolling bodies. As accelerometers do not directly measure acceleration, but instead measure forces, the simulation of *spring-accelerometers* is useful to discuss these differences. This site provides not only the simulations but also learning objectives, exercises, and resources. (E)

4. **Amusement Park Science: An AAPT Digi Kit,** C. Hall and R. Vieyra. The site https://www.compadre.org/books/AmusementPark, also hosted by the AAPT project ComPADRE, is aimed at experimenting with motion and forces in amusement park rides.It proposes several activities related to circular motion and roller coasters and discusses pedagogical aspects of the implementation. (E)

5. **Physics Experiments with Arduino and Smartphones**, G. Organtini (Undergraduate Text in Physics, Springer, in press, 2021). Focused on experimental design, data acquisition and processing at the intersection of Arduino and smartphone sensors, it discusses several experiments, especially in mechanics. Programming skills involving Python, C, and data visualization are recommended. (A)

6. **Smartphones as Mobile Minilabs in Physics,** J. Kuhn and P. Vogt, editors (Springer, in press, 2021). This edited volume features more than 70 examples from 10 years of the *iPhysicsLabs* column of *The Physics Teacher* journal. (I)

## III. HARDWARE

In Table I we list the most common smartphone sensors and their main characteristics. This Resource Letter is not focused specifically on the hardware; however, there are many important aspects that should be considered when designing an experiment with these sensors. Perhaps the most obvious fact is the existence of quantities of different physical characteristics. For example, there are sensors that measure scalar quantities, such as the ambient light sensor and the microphone, and sensors that measure vector quantities, such as the accelerometer and magnetometer. Other sensors are more difficult to classify such as the proximity sensor, on-off sensor, camera (which provides a single or a set of frames), or GPS (for geographical location). In the case of vector quantities, it is of importance to highlight that the axis refers to a system attached to the device. The orientation of the axis corresponds to the principal axis of a rectangular box.

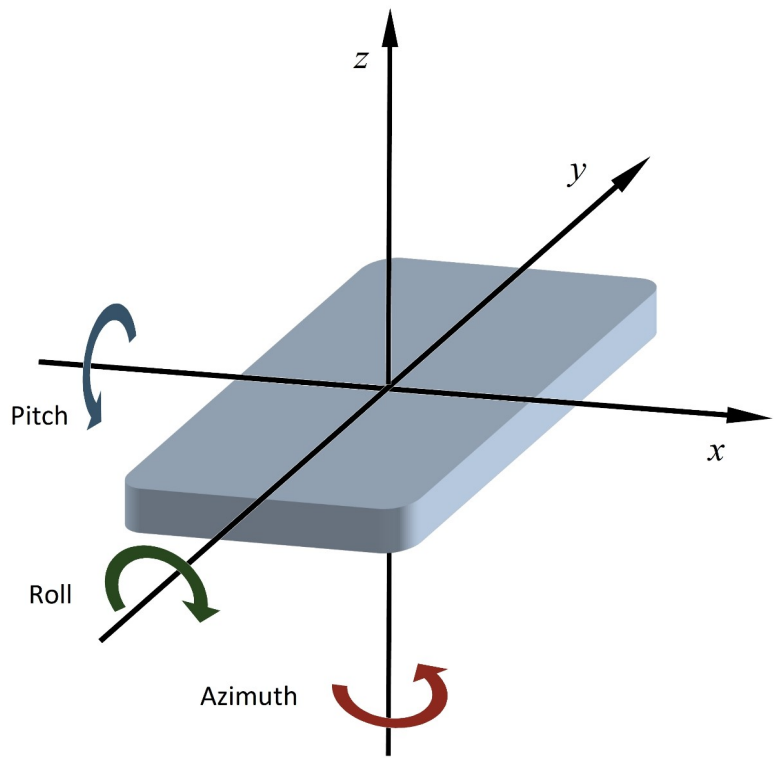

Fig. 1. Standard coordinate system of a smartphone and the orientation angles.

The precise location of the sensor inside the device could also be important in many experiments. In some cases, the location of a sensor is rather obvious: the ambient light sensor can be located simply by moving a finger back and forth above the device while watching the reading on the sensor. The magnetometer can be approximately located by scanning the device with a small magnet and watching the magnitude of the magnetic field detected. The precise location of the accelerometer is frequently needed but is difficult to find; one of the papers below describes a process for locating it.

| Table 1. Commonly Used Sensors. | | | | | |
|---|---|---|---|---|---|
| **Abbreviation** | **Sensor** | **Type** | **Quantity** | **Magnitude Provided** | **Basic purpose** |
| Ac | Accelerometer | Sensor | Vector | Apparent acceleration | Rotate screen |
| LAc | Linear accelerometers | Pseudosensor (a device that incorporates inputs from more than one sensor and after performing some calculations returns another magnitude) | Vector | Apparent acceleration minus gravitational acceleration | Detect motion |
| Gy | Gyroscope | Sensor | Vector | Angular velocity | Device's orientation |
| Mi | Microphone | Sensor | Scalar | Sound | Communication |
| Ca | Camera | Sensor | Matrix | Images | Take pictures |
| Prox | Proximeter | Sensor | Boolean | Near object detected | Disable the screen |
| Li | Light | Sensor | Scalar | Ambient light | Control display brightness |
| Pr | Pressure | Sensor | Scalar | Atmospheric pressure | Indoor and outdoor height location |
| Or | Orientation | Pseudosensor | Three angles | Around X axis (Pitch) Around Y axis (Roll) Around Z axis (Azimuth) | Device's orientation |
| GPS | GPS | Receptor | Location | Geographical location | Location |
| Ma | Magnetometer | Sensor | Vector | Local magnetic field | Device's orientation |

Below we provide a list of articles published in the physics literature that discuss the function of the sensors. In addition to the physics literature, there are also several engineering journals focused specifically on sensors, including electronics and data processing precise location of the accelerometer within the device is an important input in several experiments. An experimental method to determine the accelerometer position is also discussed here. (I, A)

available to the Arduino platform. (I)

16. "Enhance your smartphone with a Bluetooth Arduino nano board," F. Bouquet, G. Creutzer, D. Dorsel, J. Vince, and J. Bobroff, *arXiv preprint arXiv:2107.10531* (2021). The Arduino can easily be linked with the smartphone using Bluetooth, enabling a range ofnew experiments.

17. "Implementing Raspberry Pi 3 and Python in the physics laboratory," A. Martínez, C. Nieves, and A. Rúa, Phys. Teach. **59**(2), 134-135 (2021); DOI: https://doi.org/10.1119/10.0003472 . A project to measure the acceleration of falling objects is developed as an easy example of use of a low-cost Raspberry Pi 3 computer as an experimental tool. (I)

| Table 2. Useful general-purpose apps and their main characteristics. | | | | | |
|---|---|---|---|---|---|
| Name and website | Sensors | OS | Setup | Documentation and support | Other |
| Androsensor http://www.fivasim.com | Almost all | Android | Yes | Commercial adds could be annoying for the user. | A simple but powerful app to record raw data, mainly to be exported to a personal computer. Allows recording background data or while the display is turned off. |
| Phyphox https://phyphox.org | Almost all internal and external | iOS and Android | | Many packaged experiments are available, and a collaborative website offers the possibility of designing (programming) new experiments | Highly configurable app, provides the possibility of sharing screenshots, download a *csv* data file, remote controlled operation from a computer, and acquire data from external sensors. |
| Physics Toolbox Suite https://www.vieyrasoftware.net/ | Almost all | iOS and Android | | The website offers an extensive list of quick experiments and other educational apps (like Magna-AR) and games aimed at primary school and secondary students in STEM areas | A configurable app, provides the possibility of sharing screenshots or downloading a *csv* data file. |
| Arduino Science Journal https://science-journal.arduino.cc/ | Almost all internal and external | IOS and Android | | Former Science Journal or Google Science Journal. The web has a useful guided collection of science lessons suited for grades Pre-K to 12+ | A good-looking interface to collect numerical or graphical data from internal or external sensors (from Arduino or other boards) |
| Sensor Kinetics https://www.innoventions.com/ | Almost all | iOS and Android | | | A simple app to read data on the screen or to record data to be exported to a personal computer. |
| Pasco SPARKvue https://www.pasco.com/downloads/sparkvue | Internal accelerometer and many external sensors | iOS and Android | | | One of the few tools to analyze data on a mobile device. Despite being designed to work with external sensors, it also allows working with the internal accelerometer of the smartphone or tablet. |
| Magna-AR https://www.vieyrasoftware.net/ | Magnetometer | iOS and Android | | By the same developers as Physics Toolbox Suite devoted to experiment with augmented reality | Offers the possibility to visualize the magnetic field in simple setups. |

## IV. USEFUL APPS

There are many apps in the digital app stores designed to interact with the sensors. Several of them propose lesson plans in their web sites. In Table 2 we display some of them we found particularly useful and their main characteristics. We highlight Androsensor, Phyphox, and Physics Toolbox Suite for their multiplatform character, abilities to download files to the cloud, set up the sampling frequencies, documentation, and support availability.

18. "Advanced tools for smartphone-based experiments: phyphox," S. Staacks, S. Hutz, H. Heinke, and C. Stampfer, Phys. Educ. **53**(4), 045009, 6 pages (2018). DOI: https://doi.org/10.1088/1361-6552/aac05e. Phyphox was specifically designed to improve the usefulness of smartphones in a wide range of physics experiments, and this paper describes some of its features including remote access, data analysis, and web-based programmable customization. (I)

19. "Phyphox app in the physics classroom," R. Carroll and J. Lincoln, Phys. Teach. **58**(8), 606-607 (2020). DOI: https://doi.org/10.1119/10.0002393. The "technology in the classroom" column of The Physics Teacher journal reviews the main characteristics of phyphox and discusses some of the "ready-to-go" experiments proposed. (I)

## V. GENERAL EXPERIMENTS

The following references cover a broad range of physics topics and are organized in the way that introductory courses are taught.

20. "Science in your pocket," R.F. Wisman and K. Forinash, Int. J. Hands-on Sci. **1**(1), 7-14 (2008), http://www.ijhsci.info/wp-content/uploads/2008/09/IJHSCI_volume1_numero1_2008_online.pdf. This paper makes the analogy between smartphones and Swiss-army (or a multi-tool pocket) knives as useful tools for multiple everyday tasks such as measuring acceleration, analyzing sound, or even performing fast-Fourier transforms (FFT). (I)

21. "Mobile Science," R. Wisman, K. Forinash, Ubiquitous Learn. **3**(1), 21-34 (2010). DOI

meters and has uncertainties of about ±10 cm. The results are compared with those obtained using other sensors such as the accelerometer, magnetometer, and gyroscope of a smartphone, offering the possibility of several interesting discussions. (I)

25. "Time measurements with a mobile device using sound," R. F. Wisman, G. Spahn, and K. Forinash. Phys. Educ. **53**(3) 035012, 6 pages (2018). DOI: https://doi.org/10.1088/1361-6552/aaaa53. The precise determination of time intervals by means of the smartphone microphone enables experiments including a) comparing the free fall time of two balls with different initial horizontal velocity, b) obtaining the coefficient of restitution of a bouncing object, c) measuring the Doppler shift of a smartphone carried in a cyclist's pocket, and d) measuring human reaction time. (I)

26. "Using mobile-device sensors to teach students error analysis," M. Monteiro, C. Stari, C. Cabeza, and A. C. Martí, Am. J. Phys. **89**(5), 477-481 (2021). DOI: https://doi.org/10.1119/10.0002906. In contrast with most of the experiments using smartphones which use the mean value of the measurements taken by the sensors, this article focuses on the role of the fluctuations detected by the sensors. It shows how to use smartphone sensors to teach fundamental concepts in error analysis such as the physical meaning of the mean value and standard deviation and the interpretation of histograms and distributions. Applications to everyday situations are described, such as evaluating the steadiness of the device placed on a table or held in different ways and measuring the smoothness of a road. (I, A)

**VI. MECHANICS**

Since accelerometers were one of the first sensors to be repurposed for experimentation, classical mechanics is the discipline that has received most attention. In addition, in a wide range of situations the microphone can be used as a stopwatch to accurately measure time intervals between sounds. Another sensor that can be useful in classical mechanics is the ambient light sensor that can be turned into a distance sensor after suitable calibration. This section is devoted to experiments in classical mechanics (see also Refs. [20-25]).

projectile, taking into account the drag force. (I)

is also seen when a smartphone is thrown into the air while its angular velocity is recorded by the three-axis gyroscope sensor. Quantitative results are analyzed in the light of Euler's equations. (A) the acceleration vs. time, allowing a semi-qualitative analysis of the relationship with the change of momentum. (E) pendulum and the spring pendulum, is a non-typical mechanical system with two degrees of freedom. This experiment is appropriate for studying normal frequency, normal modes, and resonances. The experimental setup turns out to be very simple when using the accelerometer of a smartphone to measure horizontal and vertical acceleration. (E)

variations of the air pressure in the air track. (I)

the cup. Its sensor measures the acceleration while the device is oscillating or turning in a vertical plane. The qualitative comparison of the radial and tangential components of the acceleration is used to explain the functioning of the device. (E)

MATLAB software in real time, facilitating graphics visualization. (I)

## VII. WAVES AND SOUND

Oscillations and waves can be studied using smartphone microphones and speakers. Several experiments typically performed using stopwatches have been redesigned to use smartphones, improving their precision. These experiments address concepts such as stationary waves, harmonics, resonance, the speed of sound in different media. In addition, the high processing speed of built-in CPUs allows real-time fast Fourier transform (FFT) analysis to identify the main frequencies of a sound wave, measure the resonant frequency of a system, or *visualize* a melody in a spectrogram. It is worth noting that most of the experiments proposed in this section are very economical (see also Refs. [21, 22, 25]). Also note that in some modern smartphones the audio port is replaced by Bluetooth alternatives. Ref. [16] describes how external sensors can be connected via Bluetooth.

for measuring the speed of sound in air is revisited. A smartphone is used as a signal generator, and a glass pipe is used as a resonator, with one end immersed in water to change the length and find the nodes of sound waves. (E)

## VIII. LIGHT AND OPTICS

The integrated camera, the liquid crystal display, and the ambient light sensor are especially suited to perform experiments involving light and optics. The experiments cover a wide range of topics from geometrical optics to human vision or shadowgraphy. The use of smartphones as simple but powerful microscopes enables a range of experiments.

an RGB LED, as a way of improving students' understanding of color mixing. (E)

## IX. ELECTRICITY AND MAGNETISM

Thanks to the magnetic field sensor, several experiments have been designed to measure magnetic fields produced by currents or magnets. Most of these experiments are easy and inexpensive to perform. In contrast, experiments involving measurements of voltages or currents are possible, using the microphone input, transforming the smartphone into a portable oscilloscope. These are limited due to the danger of damaging the mobile device by connecting a large voltage. A safer alternative is to send signals to the smartphone via Bluetooth from devices such as Arduino BLE, as described in Ref. 16.

111. "Magnetic field sensor," N. Silva, Phys. Teach. **50**(6), 372-373 (2012). DOI: https://doi.org/10.1119/1.4745697. One of the most direct uses of the smartphone magnetometer measure the magnetic field of current-carrying coil as a function of the number of turns of the coil. (E)

112. "Demonstrations of magnetic phenomena: Measuring the air permeability using tablets," V. O. M. Lara, D. F. Amaral, D. Faria, and L. P. Vieira, Phys. Educ. **49**(6), 658 (2014). DOI: https://doi.org/10.1088/0031-9120/49/6/658. This paper describes a simple and economical experiment to find the dependence of a magnetic field on the distance to a permanent magnet or a coil carrying an electric current. Air permeability can also be accurately obtained. (E)

113. "Measurement of the magnetic field of small magnets with a smartphone: A very economical laboratory practice for introductory physics courses," E. Arribas, I. Escobar, C. P. Suarez, A. Najera, and A. Beléndez, Eur. J. Phys. **36**(6), 065002, 11 pages (2015). DOI: https://doi.org/10.1088/0143-0807/36/6/065002. The dependence of the magnetic field of small refrigerator magnets with distance is measured in this inexpensive experiment. It also provides a good opportunity to discuss least-square fitting and uncertainties. (E)

114. "Linear quadrupole magnetic field measured with a smartphone," E. Arribas, I. Escobar, R. Ramirez-Vazquez, C. del P. Suarez Rodriguez, J. Gonzalez-Rubio, and A. Belendez, Phys. Teach. **58**(3), 182-185 (2020). DOI: https://doi.org/10.1119/1.5145411. An experiment is described that uses widely available materials to find the dependence on distance of the magnetic field produced by a quadrupolar arrangement of permanent magnets. (E) record the variations in electrical potential due to cardiac activity in the body of a volunteer. Especially suitable for Introductory Physics for Life Sciences, it involves the concepts of electric potential, oscillating dipoles, current, electrodes, and impedance. (It is not intended for medical use.) (I)

121. "Simple determination of Curie temperature using a smartphone magnetometer," B. W. Nuryadin and R. Rusman, Phys. Teach. **57**(6), 422-423 (2019). DOI: https://doi.org/10.1119/1.5124290. When the temperature of a ferromagnetic sample is raised above the Curie point, it becomes a paramagnet. In this experiment, the Curie temperature of a permanent magnet is analyzed quantitatively using a smartphone magnetometer. This is one the few experiments involving statistical concepts such as phase transitions and mean field approach. (I)

122. "Electric circuits as seen by thermal imaging cameras," P. Kácovský, Phys. Teach. **57**(9), 597-599 (2019). DOI: https://doi.org/10.1119/1.5135785. Thermal cameras designed to be attached to smartphones provide the opportunity to perform several interesting experiments and demonstrations. Pointing the thermal camera at a DC circuit allows visualization of thermoelectric processes involving Kirchhoff's or Ohm's laws. (E)

123. "An indirect measurement of the speed of light in a general physics laboratory," E. Arribas, I. Escobar, R. Ramirez-Vazquez, T. Franco and A. Belendez, J. King Saud Univ. Sci. **32**(6), 2797-2802 (2020). DOI: https://doi.org/10.1016/j.jksus.2020.06.017. The speed of light is obtained from measurements of the permeability and permittivity of free space. These experiments require a parallel-plate capacitor, coils, multimeters, and a smartphone used as a magnetometer. (I)

124. "A semester-long study of magnetic fields using smartphones to engage non-physics majors," S. A. Hootman and C. Pickett, Phys. Teach. **59**(2), 108-110 (2021). DOI: https://doi.org/10.1119/10.0003463. Students take part in a project to measure the magnetic field at different locations on their university campus. This activity is especially suited for life

sciences students since it also discusses health-related and geophysical issues. (E)

## X. MATTER, FLUIDS, THERMAL PHYSICS

Since smartphone batteries dissipate a considerable amount of energy, temperature sensors are not usually included in these devices. (But see commercial attachments at https://thermodo.com.) Thus, thermodynamics experiments for smartphones are rare.

Nonetheless, pressure sensors are available in several smartphone models and experiments in a variety of topics have been discussed. Other experiments have been proposed using Arduino-based sensors, and other ingenious setups are based on the cellular radio-frequency receptor.

125. "Exploring the atmosphere using smartphones," M. Monteiro, P. Vogt, C. Stari, C. Cabeza, and A. C. Martí, Phys. Teach. **54**(5), 308-309 (2016). DOI: https://doi.org/10.1119/1.4947163. Altitude and pressure of the lowest layer of the atmosphere are measured using a smartphone mounted on a quadcopter. Measurements can be compared with other simple approximations: isothermal, constant density, and the International Standard Atmosphere. (I)

126. "Using smartphone pressure sensors to measure vertical velocities of elevators, stairways, and drones," M. Monteiro and A. C. Martí, Phys. Educ. **52**(1), 015010, 11 pages (2017). DOI: https://doi.org/10.1088/1361-6552/52/1/015010. Pressure sensors, or barometers, although not present in many smartphones, are very useful to measure heights and vertical velocities in several contexts allowing a wide variety of indoor and outdoor experiments. (I)

127. "Using a cell phone to investigate the skin depth effect in salt water," J. Rayner, Phys. Teach. **55**(2), 83-86 (2017). DOI: https://doi.org/10.1119/1.4974118. This article describes a simple demonstration of skin depth using only a small container filled with salty water and a waterproof smartphone as a sensor. Several interesting activities are proposed; however, the experimental conditions are not fully controlled, and there is no direct calibration of the relationship between the amplitude of the field and the cellphone's measurements. (I)

128. "Analyzing Stevin's law with the smartphone barometer," S. Macchia, Phys. Teach. **54**(6),

Tornaría, and A. C. Martí, Eur. J. Phys. **41**(3), 035005, 8 pages (2020). DOI: https://doi.org/10.1088/1361-6404/ab7f81. The shape of the surface of a fluid in a rotating container is obtained using a gyroscope sensor and video analysis, and its dependence on the angular velocity is analyzed. (I)

134. "Surface tension measured with Arduino," A. M. B. Goncalves, W. P. S. Freitas, D. D. Reis, C. R. Cena, D. C. B. Alves, and D. F. Bozano, Phys. Teach. **57**(9), 640-641 (2019). DOI: https://doi.org/10.1119/1.5135800. Through measurements of the weight of a water droplet suspended at the tip of a pipette and video analyses, the surface tension between air and water is obtained. (E)

## XI. MODERN PHYSICS

Although smartphones are not especially well-suited to performing experiments in modern physics, several creative experiments have been put forward, including ones that enable experiments on radioactivity or requiring particle detectors. The advantage of smartphones in these contexts is clear.

135. "iRadioactivity - Possibilities and limitations for using smartphones and tablet PCs as radioactive counters," J. Kuhn, A. Molz, S. Gröber, and J. Frübis, Phys. Teach. **52**(6), 351-356 (2014). DOI: https://doi.org/10.1119/1.4893089. By covering the camera with a black tape, it is possible to experiment with radioactivity in an educational setting. The experiments proposed focus on the inverse-square law for distance, absorption laws, and decay of meta-stable isotopes. Although camera sensors are sensitive to radiation, the characteristics strongly depend on the specific sensor. Since general sensors are not sensitive enough to experiment with background radiation, a standard radioactivity source, like those present in many educational laboratories, is needed. A Geiger counter is advisable for comparison.

136. "Using smartphones and tablet PCs for $\beta^-$-spectroscopy in an educational experimental setup," S. Gröber, A. Molz, and J. Kuhn, Eur. J. Phys. **35**(6), 065001 (2014). DOI:

https://doi.org/10.1088/0143-0807/35/6/065001. A β–spectrometer can be built using a PC or a mobile device, a magnetometer, a simple electric circuit, and a commercial $^{90}$Sr/Y source. A Geiger-Müller tube is also advisable for comparison. As in the previous reference, the app, RadioactivityCounter, measures radioactive radiation with the camera sensor. (A)

137.    "A low-cost computer-controlled Arduino-based educational laboratory system for teaching the fundamentals of photovoltaic cells," K. Zachariadou, K. Yiasemides and N. Trougkakos, Eur. J. Phys. **33**, 1599-1610 (2012). DOI: https://doi.org/10.1088/0143-0807/33/6/1599. This paper describes an experiment to learn about semiconductor physics.  Note that there are currently more suitable alternatives to be used than the apps proposed here. (I)

138.    "The desktop muon detector: A simple, physics-motivated machine- and electronics-shop project for university students," S. N. Axani, J. M. Conrad, and C. Kirby, Am. J. Phys. **85**, 948 (2017).  DOI: https://doi.org/10.1119/1.5003806. For less than $100 it is possible for undergraduates to construct a muon detector using a silicon photomultiplier to detect scintillation light from particles and an Arduino board for data acquisition. Its construction develops machine-shop, electronics-shop, and programming skills. (A).

## XII. ASTRONOMY

Experiments related to astronomy make use of the camera as an observational tool or that simulate astronomical concepts such as parallax, planetary transits and occultations. The ambient light sensor also plays a key role in several of these experiments.

139.    "Using smartphone camera technology to explore stellar parallax: Method, results, and reactions," M. T. Fitzgerald, D. H. McKinnon, L. Danaia, S. Woodward, Astron. Educ. Rev. **10**(1), 010108, 8 pages (2011). DOI: http://dx.doi.org/10.3847/AER2011028. A simple and practical method is proposed to introduce the parallax technique to determine distances in a scaled system that uses data from a smartphone's camera, simulating the way astronomers measure some stellar distances. (I)

## XIII. PHYSICS EDUCATION RESEARCH

Although the use of smartphones in physics laboratories is a recent introduction, researchers have already started to explore their impact on student learning.

153. "A guide for incorporating e-teaching of physics in a post-COVID world," D. J. O'Brien, Am. J. Phys. **89**(4), 403-412 (2021). DOI: https://doi.org/10.1119/10.0002437. In a paper discussing many aspects of e-learning, the advantages of smartphones are explored as educational and experimental tools. Nearly 80 examples of smartphone-based labs and home introductory physics experiments are provided and sorted by subject. (I)

154. "Lessons from transforming second-year honors physics lab," D. Doucette, B. D'Urso, and C. Singh, Am. J. Phys. **88**, 838 (2020). DOI: https://doi.org/10.1119/10.0001641. Focused on the introduction of new technology, this paper proposes three laboratory modules encouraging students' initiative in opposition to a "black box" philosophy. (I)

## XIV. CITIZEN SCIENCE

Mobile-device sensors are especially suited to a broad range of experiments that enable public participation in identifying research questions, collecting and analyzing data, and understanding environmental problems.

155. **iSPEX (http://ispex.nl/en/). Air pollution.** A citizen science project developed in the Netherlands in 2012 to measure atmospheric aerosols, based on a free app and a low-cost add-on for smartphone cameras. The instrument is based on that of the Spectropolarimeter for Planetary EXploration (SPEX). The citizen scientists scan the sky while the phone's camera takes pictures through the add-on, recording the spectrum and the linear polarization of the sunlight that is scattered by suspended dust particles, making a map of contaminants. (A)

156. **CrowdMag (https://www.ngdc.noaa.gov/geomag/crowdmag.shtml). Geomagnetic field.** A citizen science project developed by the geomagnetism group of NOAA's National Centers for Environmental Information (NCEI). The CrowdMag app uses the built-in magnetometer of the phone of volunteers around the world to send magnetic field data to the server of the research group. The main goal is to use the citizen generated

information to fill in gaps in measurements of Earth's magnetic field. (I)

157.     **MyShake (https://myshake.berkeley.edu/). Earthquake detection.** A citizen science app developed by researchers from the University of California, Berkeley, that takes advantage of the ubiquity of smartphones, their built-in GPS, and tri-axis accelerometers to detect earthquakes. The aim is to build a global earthquake early warning network that does not replace the standard networks but could be useful in some seismically active places. More information about the technical development of the project can be found in the research paper, "MyShake: A smartphone seismic network for earthquake early warning and beyond," Q. Kong, R. M. Allen, and L. Schreierand and Y. Kwon, Sci. Adv. 2(2), e1501055, 1-8 (2016). DOI: https://doi.org/10.1126/sciadv.1501055. (A)

158.     **DECO (https://wipac.wisc.edu/deco), Cosmic Rays.** The Distributed Cosmic-ray Observatory (DECO) is a citizen science project led by Justin Vandenbroucke at the University of Wisconsin–Madison supported by the National Science Foundation and the American Physical Society, among others. The DECO app enables users around the world to detect cosmic rays and other energetic particles using the camera sensors of their smartphones and tablets. The recorded events are automatically uploaded to a central database. In addition to detecting particle events, users can analyze the data produced by their own or other users' phones. This work is described in the paper: "Particle identification in camera image sensors using computer vision," Astropart. Phys. **104**, 42-53 (2019). DOI: https://doi.org/10.1016/j.astropartphys.2018.08.009. (A)

159.     "Gravimetric detection of Earth's rotation using crowdsourced smartphone observations," S.F. Odenwald and C. M. Bailey, IEEE Access **7**, 148131-148141 (2019). DOI: https://doi.org/10.1109/ACCESS.2019.2940901. The rotation of the Earth is detectable by means of crowdsourcing measurements of the slight changes in local gravity that can be detected by smartphone accelerometers. (I)

160.     "Smartphone science: Apps test and track infectious diseases," S. Ravindran, Nature,

## XV. FINAL REMARKS

In the last few years, the use of smartphones and their sensors has had a great impact on physics teaching. In this Resource Letter we have concentrated on the application of sensors available in a wide range of mobile devices, setting aside other uses such as access to information on the internet, preparation of presentations, exchange of ideas between peers or with the teacher, and more.

It is our view that the most important characteristic in the use of these devices and their sensors is their adaptability to a wide range of applications. While sensors are most easily applied to experiments in mechanics or waves, they can be used in every possible field. They enable experiments illustrating elementary concepts, but also can be used in advanced projects that require considerable mathematical and experimental skills. While accelerometers are most commonly used, the camera, microphone, and GPS locator are also frequently used, even though they do not fall within the usual definition of a sensor. Some experiments are based on measuring scalar quantities, but others involve vectors, logical variables, or matrices. The software used, which works under different operating systems, ranges from very elaborate programs to simple applications, free or purchased. In some cases, data are transferred to a computer for processing; in others they are processed on the device itself. Many of the experiments proposed are carried out in laboratories; however, others are carried out outdoors, in amusement parks, or on means of transport. Some require laboratory materials or equipment, while others can be carried out with material available in many homes. There are many advantages to the use of these sensors. Their wide availability is an important factor. Even in socially disadvantaged contexts there is usually access to mobile devices at no additional cost, while access to traditional educational materials may be limited or nonexistent. Their portability and the convenience of having multiple sensors in one device that communicate directly with each other is another great advantage, allowing measurements and connections that would otherwise be difficult to achieve.

It is worth asking whether using the sensors in mobile devices represents a significative advance or only a marginal one. The same question could be asked about the use of smartphones in other areas, and we believe the answers are essentially the same: access to a set of resources in a single device, at an affordable price, and in a compact size, is, by itself, a great advantage. The definitive answer will be provided by users themselves when they choose one alternative or another.

There are many unknown factors to be faced in the coming years. Mobile devices have incorporated sensors only over the last decade. We do not know whether manufacturers will continue to include new sensors, or whether their calculation capacity will continue to increase at the same rate as in recent years.

Perhaps future advances will improve the ease of use of software and data processing. Physics education research also has much to say, indicating whether future change will be significant or merely marginal, and in what aspects improved learning outcomes will be noted. In any case, what we can state is that these devices and sensors are here to stay. As with all new tools or technology that appear, the best attitude seems to be to analyze their contributions and use them critically and reflectively.

**ACKNOWLEDGEMENTS.** The authors are very grateful to the anonymous reviewers for their careful reading and thoughtful comments. We appreciate their suggestions and have included many of them in this Resource Letter.